# A predictive TCAD module for total-ionizing-dose effects based on partially coupled defect kinetics

**Bo Chen[1,2,†], Yu Song[2,3,†,*], Xiaowen Liang [2,3], Henglei Chen[1,2,*], Xiaolong Li[2,3], Lin Wen[2,3], Qi Guo[2,3] and Yudong Li [2,3,*]**

[1] School of Physical Science and Technology, Xinjiang University, Urumqi 830017, China
[2] Xinjiang Key Laboratory of Extreme Environment Electronics, Xinjiang Technical Institute of Physics and Chemistry, Chinese Academy of Sciences, Urumqi 830011, China
[3] University of Chinese Academy of Sciences, Beijing 100049, China
† These authors contribute equally to this work.
*Author to whom any correspondence should be addressed.
**E-mail:** songyu@ms.xjb.ac.cn; chl@xju.edu.cn; lydong@ms.xjb.ac.cn

## Abstract

Predicting total-ionizing-dose (TID) effects with technology computer-aided design (TCAD) can reduce irradiation time and support mechanism identification. Here, we integrate our previously developed analytical model of partially coupled oxide-trap and interface-trap kinetics into Sentaurus and test the resulting predictive module using gate-controlled lateral PNP (GLPNP) transistors fabricated with a commercial integrated-circuit process. First, we demonstrate that, at a relatively high dose rate of 1 rad/s, a single set of kinetic parameters jointly fitted to the two defect concentrations over 1–10 krad quantitatively reproduces the measured emitter-current–gate-voltage (DS) curves in this range and predicts the full DS curve at the held-out dose of 60 krad. In these workflows, the device parameters remain unchanged after calibration to the pre-irradiation I–V data. Then, we use the calculated concentrations as a baseline to identify additional trap properties needed to quantitatively reproduce the remaining electrical responses. It is found that, at lower dose rates, the DS curve shapes require nonuniform interface-state energy distributions, represented here by Gaussian functions. Moreover, the base-current–gate-voltage (GS) curves at 1 and 0.1 rad/s require an effective hole-capture cross section that varies with dose and dose rate. Finally, we show that, among the uncoupled, fully coupled, and partially coupled kinetics models, only the last simultaneously reproduces the nonlinear dose dependence of both defect concentrations required by the prediction. Our TCAD module thus demonstrates prediction from smaller to larger total dose under fixed irradiation conditions and identifies additional trap information needed for quantitative device simulation.

## 1. Introduction

Total ionizing dose (TID) effects remain a major reliability concern in space, accelerator, and nuclear electronics [1], [2]. This is because, although gate-oxide scaling has reduced the sensitivity of core CMOS transistors, modern technologies still contain thicker isolation, spacer, and buried oxides, while bipolar and power devices retain radiation-sensitive $SiO_2$/Si regions [3]–[9]. Ground irradiation is indispensable for radiation qualification and for identifying TID mechanisms, but only a limited number of doses and conditions can be tested. Therefore, TCAD simulation of device damage is valuable for reproducing measured degradation curves, identifying mechanisms, and supporting radiation-hard design. Moreover, using TCAD simulation to predict responses at larger doses would further reduce irradiation time and cost [10]-[14]. Both functions require a model that calculates defect concentrations at an unmeasured dose and supplies them to the device solver.

Modern device simulators can calculate TID-induced changes in electrostatics, transport, and recombination when the oxide-trap density ($N_{ot}$), interface-trap density ($N_{it}$), interface-state energy distribution, and capture cross sections are specified [10], [11], [13]–[16]. However, determining how these quantities evolve with the irradiation dose remains more difficult.

Depending on the available measurements, TID simulations commonly use one or more of three strategies: 1) $N_{ot}$ and $N_{it}$ are extracted at individual doses and inserted into TCAD; 2) effective trap parameters are assumed or calibrated to reproduce measured I–V curves when the two concentrations cannot be separated experimentally; or 3) the measured or extracted quantities are fitted with simple empirical dose laws [10]–[14], [16], [17]. These approaches can reproduce measured responses under the calibrated conditions. However, calibrating concentrations independently at each dose does not specify their values at an unmeasured dose, and simple empirical laws may extrapolate poorly. The Sentaurus manual example [15] examined here illustrates the latter case. It will be shown that with its fixed parameters, the *uncoupled* models of $N_{ot}$ and $N_{it}$ produce an interface-trap-dominated rightward shift of the NMOS transfer curve, whereas comparable irradiated MOS structures show an oxide-charge-dominated leftward shift [3], [18]–[20].

Our previous work [21] developed an analytical model for the dose dependence of $N_{ot}$ and $N_{it}$ that can support TCAD simulations. It combines *retarded* oxide-trap generation in ionizing-irradiated amorphous $SiO_2$ with *partial* coupling between oxide-trap passivation and interface-trap formation [2], [7], [21]–[27]. Here, partial coupling means that only a fraction of the generated oxide traps undergo hydrogen-mediated passivation that produces interface traps. Physically, dispersive transport and trapping of radiation-induced holes by oxygen vacancies of various configurations lead to retarded oxide-trap generation [24], [28]. On the other hand, hydrogen interactions, proton generation and transport, and the structure of interface defects provide the microscopic basis for the coupling [29]–[35], while reversible hydrogen reactions and space charge limit its extent [21], [25]–[27]. Comparisons with experiments showed that this partially coupled model reproduces sublinear $N_{ot}$ buildup at relatively high dose rates, nonmonotonic $N_{ot}$ evolution at very low dose rates, and an approximately fractional-power-law dose dependence of $N_{it}$ at both high and low dose rates [21], [25]. Related analytical kinetics have also supported predictive modeling of post-irradiation annealing in time-delay-integration (TDI) CMOS image sensors [36]. However, whether concentration extrapolation can predict device I–V curves had not been established. This is the first motivation of the present work. Moreover, experiments have shown that interface-state energy distributions and capture properties can change with irradiation [37]–[40]. A second motivation is therefore to determine when defect concentrations alone are insufficient for device simulation and what additional trap information is needed.

Gate-controlled lateral PNP (GLPNP) transistors allow us to test both the predicted concentrations and the additional trap information needed to quantitatively calculate electrical responses. This is because, their emitter-current–gate-voltage ($I_e$–$V_g$, DS) characteristic is mainly controlled by electrostatics and is therefore sensitive to defect concentrations. On the other hand, their base-current–gate-voltage ($I_b$–$V_g$, GS) characteristic is strongly sensitive to interface-state-assisted surface recombination [41]–[45]. More generally, microscopic calculations show that defect-state structure can strongly influence electron and hole capture and recombination [46].

In this work, we integrate the analytical partially coupled defect kinetics model from our previous work [21] into Sentaurus as a predictive TID simulation module. First, we calibrate the device parameters against the pre-irradiation GLPNP DS curve. Importantly, we demonstrate that, at relatively high dose rate of 1 rad/s, one set of kinetic parameters fitted to the 1–10-krad defect-concentration data can reproduce the DS curves at multiple doses, and predicts the full 60-krad curve. Second, we show that calculations with uniform energy distributions and fixed capture cross sections identify when concentrations alone are insufficient. It is found that lower-dose-rate DS curves require a nonuniform interface-state energy distribution, represented here by Gaussian functions, and the GS curves examined here additionally require effective capture cross sections that vary with dose and dose rate. In addition, the NMOS benchmark and comparisons with uncoupled [15] and fully coupled [47] defect kinetics models support the use of our partially coupled defect kinetics model to develop predictive TCAD in the dose domain.

## 2. Development of predictive TID TCAD module in the dose domain

### 2.1. Module implementation and Sentaurus manual benchmark

Figures **1(a)** and **(b)** compare the Sentaurus manual workflow with our implementation. Both use Sentaurus Structure Editor (SDE) to construct the device, calculate defect concentrations as continuous functions of total dose, and use them in SDevice to calculate the electrical response at different total doses. The difference is that, we replace the manual defect-concentration formulas with ours. Python is used to evaluate the two defect concentrations at each requested dose with our model, and Sentaurus Workbench (SWB) passes $N_{ot}$(D) and $N_{it}$(D) to SDevice. Both quantities are applied at the Si/SiO$_2$ interface as areal densities (cm$^{-2}$); $N_{it}$ is represented by uniformly distributed donor- and acceptor-like states. **Figure 1(c)** shows the integrated Python module.

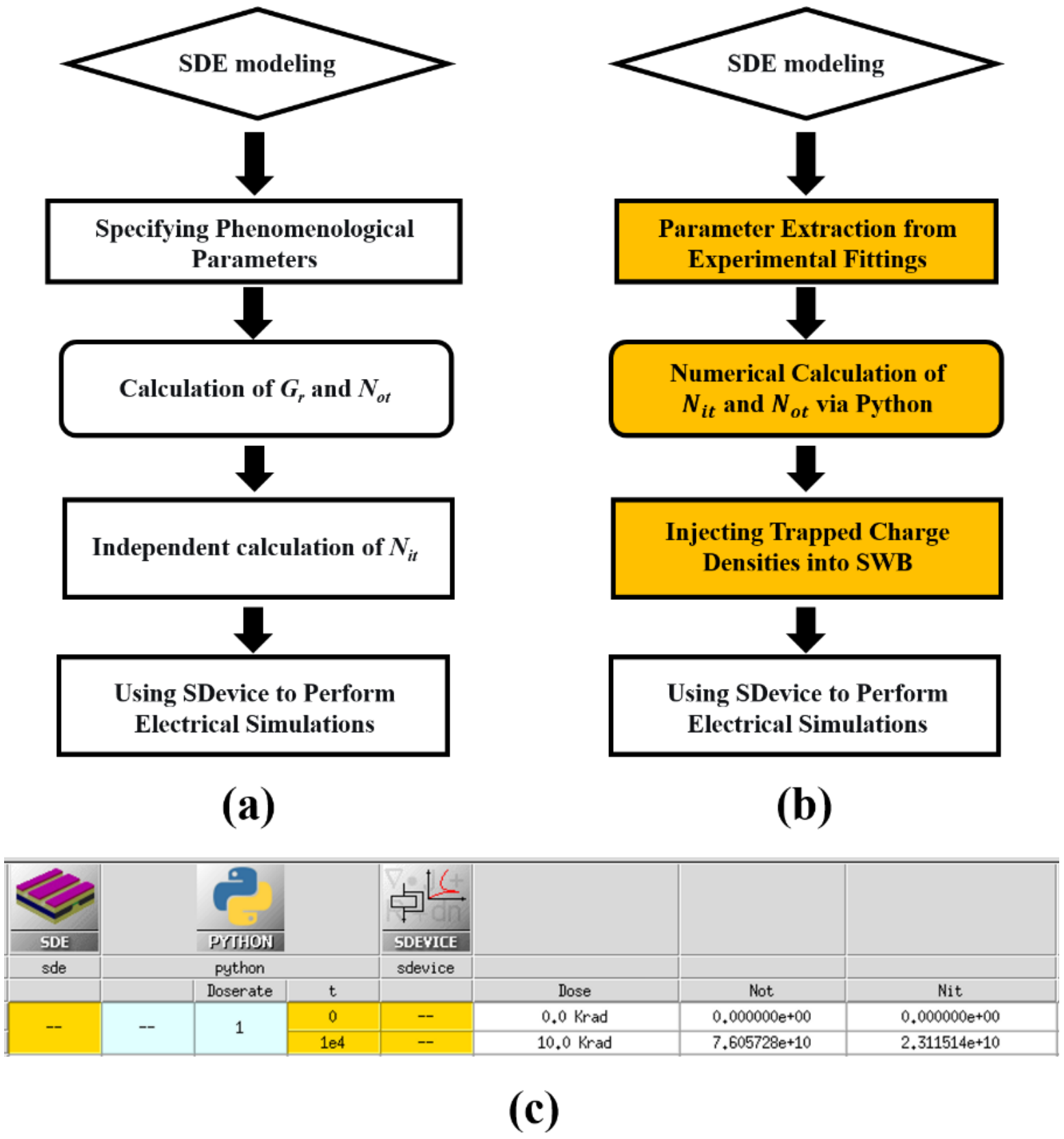

**Figure 1.** TID simulation workflows. (a) Sentaurus manual using uncoupled dose laws and fixed parameters. (b) Our Python implementation of the partially coupled model, supplying $N_{ot}$(D) and $N_{it}$(D) to SDevice through SWB. (c) Overview of our Python module integrated into Sentaurus.

The manual benchmark uses uncoupled $N_{ot}$ and $N_{it}$ defect-kinetics models with fixed parameters. The radiation-generated electron–hole-pair rate is $G_r = g_0\dot{D}Y(F)$, with $Y(F) = \left[\frac{(F+E_0)}{(F+E_1)}\right]^m$ [15]. The fixed parameter set is $g_0 = 7.6 \times 10^{12}$ cm$^{-3}$ rad$^{-1}$, $E_0$ =0.1 V/cm, $E_1 = 1.35 \times 10^6$ V/cm, and $m$=0.9. Here, $g_0$ is the pair yield per unit absorbed dose and volume, $E_0$ and $E_1$ set the empirical field scales, and $m$ controls the field dependence. The oxide-trap density $N_{ot}$ (denoted $N_b$ in the manual) is calculated from $G_r$ using

$$N_{ot}=(1-\alpha_{ot})G_r t \tag{1a}$$

On the other hand, the interface-trap density $N_{it}$ is calculated *independently* as a sublinear function of dose,

$$N_{it}=(1-\alpha_{it})a_{it}D^{b_{it}} \tag{1b}$$

where $\alpha_{ot}$ and $\alpha_{it}$ are normally set to zero and therefore do not affect the dose dependence. The parameters $a_{\text{it}}$ and $b_{\text{it}}$ in Eq. (1b) are empirical coefficients obtained by fitting experimental data. In the manual formulation, they are treated as device-type-dependent constants. For NMOS devices, $a_{\text{it}}$=1.1 × $10^6$ and $b_{\text{it}}$=0.95 [17].

Following the manual workflow, we simulated an NMOS irradiated at 5×$10^4$ rad/s for 100 s (the total dose is 5 Mrad). As shown by the red dashed curve in **Fig. 2**, the transfer curve shifts to the right, as in the manual example. However, this is opposite to the oxide-charge-dominated negative threshold-voltage shift commonly observed experimentally [3], [18]–[20]. Repeating the calculation at 100 rad/s for 100 s (total dose is 10 krad) also gives a rightward shift (blue dashed curve in **Fig. 2**). To test if this is due to too few positively charged oxide traps relative to negatively charged interface traps, we kept $N_{it}$ fixed and multiplied $N_{ot}$ by $10^3$. As expected, this reverses the shift (green solid curve in **Fig. 2**).

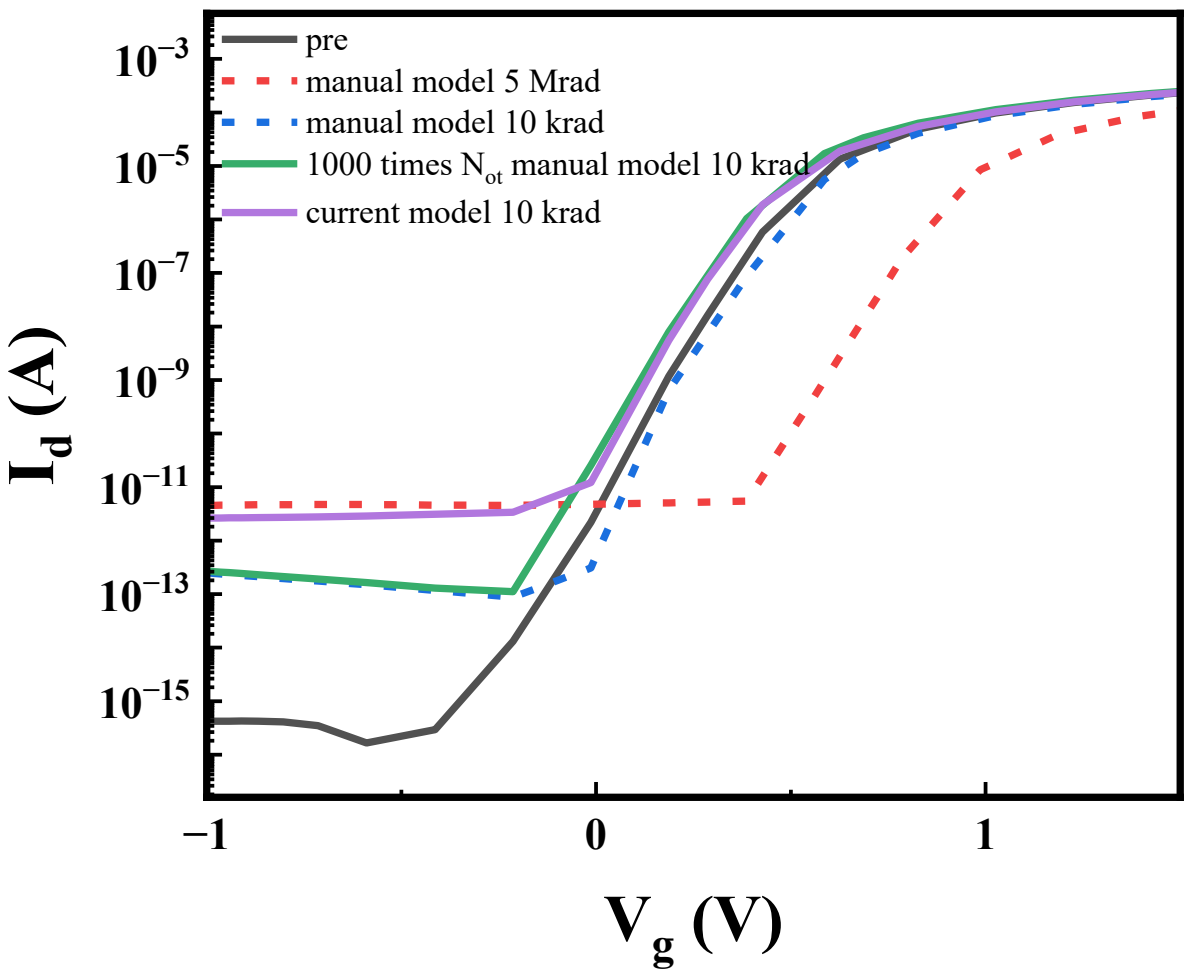

**Figure 2.** NMOS transfer curves calculated using different defect concentrations. The manual uncoupled formulas give rightward shifts at 5 Mrad and 10 krad (red and blue dashed curves). Multiplying $N_{ot}$ by $10^3$ reverses the 10-krad shift (green solid curve). Replacing the manual formulas with the partially coupled model also restores a leftward shift, without this multiplication (magenta solid curve; see section 2.3).

With the device structure and transport solver unchanged, these tests show that the ratio between $N_{ot}$ and $N_{it}$ controls the shift direction. Inappropriate concentrations can therefore give the wrong sign of the threshold-voltage shift. Below we show that our partially coupled kinetic model provides a more appropriate description.

### 2.2. Partially-coupled defect kinetics model and Python integration

The partially coupled defect kinetics model developed in our previous work [21] calculates the total areal densities of oxide traps and interface traps based on the physics of dispersive hole trapping and reversible hydrogen reactions. The two concentrations are continuous functions of total dose $D = qt$, where $q$ is the dose rate:

$$N_{ot} = (1\text{-}\lambda_c)\kappa_g D^{\beta} + \lambda_c \kappa_g D_c^{\beta} e^{-\frac{D}{D_c}} \beta (\text{-}1)^{\text{-}\beta} \Gamma(D) \tag{2a}$$

$$N_{it} = \lambda_c \kappa_g D^{\beta} \text{-} \lambda_c \kappa_g D_c^{\beta} e^{-\frac{D}{D_c}} \beta (\text{-}1)^{\text{-}\beta} \Gamma(D) \tag{2b}$$

Here, $\Gamma(D) \equiv \Gamma(\beta, 0, -D/D_c)$ denotes the generalized incomplete gamma function. The four parameters describe different aspects of defect kinetics: $\kappa_g$ is the effective oxide-trap generation coefficient, while $\beta < 1$ describes retarded generation associated with dispersive hole trapping in amorphous $SiO_2$; $D_c$ sets the characteristic dose of the hydrogen-mediated coupling process, and $q/D_c$ is the corresponding rate constant; $\lambda_c$ is the fraction of generated oxide traps that undergo hydrogen-mediated passivation associated with interface-trap formation. The partial coupling is reflected by $\lambda_c < 1$. Thus, the model links the evolution of $N_{ot}$ and $N_{it}$ through hydrogen chemistry [2], [7], [21], [22], [25], [26]; it does not assume direct conversion of oxide traps into interface traps. It should be indicated that, both Eqs. (1) and (2) are continuous in dose; the distinguishing feature of Eq. (2) is the partial coupling between the two concentrations. Allowing the empirical coefficients in Eq. (1) to vary does not introduce this coupling.

As demonstrated in Fig. 1(b), for each sample and dose-rate condition, all measured $N_{ot}$ and $N_{it}$ values as functions of total dose are fitted simultaneously by Eq. (2) in the dose domain, with one parameter set $\kappa_g, \beta, D_c, \lambda_c$ for both $N_{ot}$ and $N_{it}$. This differs from fitting only $N_{it}$ as a function of dose rate at a fixed total dose. This simultaneous fit is referred to below as the joint fit. The parameter values can differ between samples and irradiation conditions. Then, Python evaluates $N_{ot}$ and $N_{it}$ through Eqs. (2a) and (2b) at any requested dose and exports the areal densities through SWB. Device geometry, doping, transport, recombination, and contact models remain in the native solver.

### 2.3. Correction of the NMOS shift direction

To test the predictive ability of our module, we first consider the Sentaurus manual benchmark. For the 100-rad/s NMOS case, we fitted the corresponding defect data reported in Ref. [3] with Eq. (2) and used the calculated concentrations in the TCAD simulation of the NMOS structure. As shown by the magenta solid curve in **Fig. 2**, the simulated transfer curve now shifts to the left. This result shows that the module based on partially coupled defect kinetics gives the expected *qualitative* shift direction. As discussed above, this qualitative correction from right-shift to left-shift follows from the correct ratio between $N_{ot}$ and $N_{it}$ calculated by our model. Below, we will test whether the module can *quantitatively* reproduce I–V curves within the dose fitting interval and predict a curve at a larger dose reserved for validation.

## 3. Predictive TID TCAD in the dose domain and experimental validation

### 3.1. Experimental setup: GLPNP devices and irradiation measurements

GLPNP bipolar junction transistors (BJTs) provide two complementary electrical measurements in the same device. The emitter-current–gate-voltage ($I_e$–$V_g$, DS) characteristic is primarily MOS-like and electrostatic, whereas the base-current–gate-voltage ($I_b$–$V_g$, GS) characteristic is strongly influenced by interface-state-assisted surface recombination. Gate-controlled bipolar methods have long been used to separate $N_{ot}$ and $N_{it}$ as functions of total dose for different irradiation conditions, such as dose rates [41]–[45].

The devices were fabricated at the National Laboratory of Analog ICs, China, using the same process as that used for its commercial integrated circuits. A cross-sectional schematic of the GLPNP transistor is shown in **Fig. 3**. All devices were obtained from the same production lot.

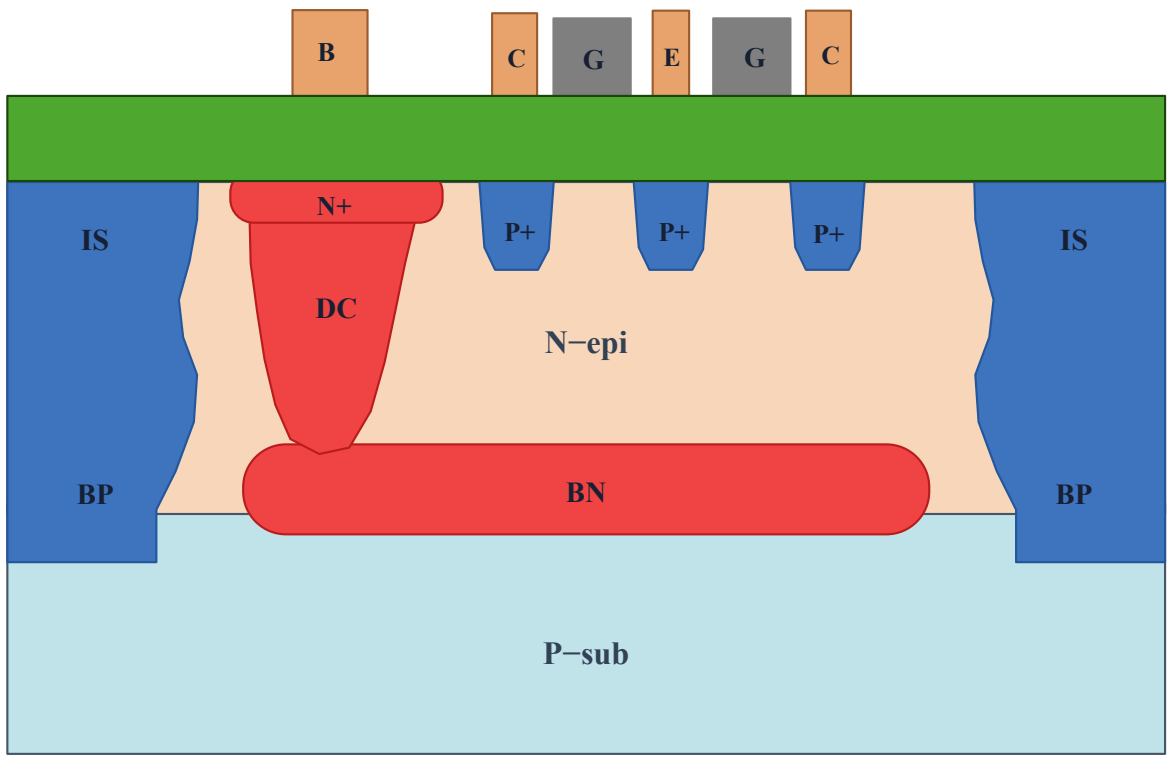

**Figure 3.** Cross-sectional schematic of the GLPNP structure used for the irradiation measurements and TCAD pre-irradiation calibration/post-irradiation prediction.

The GLPNP geometry in **Fig. 3** was implemented in SDE and held fixed across irradiation conditions. Key dimensions are a 0.75-μm oxide, a 14-μm base width, a 22-μm base–collector spacing, a 12-μm N-epi/base vertical thickness, a 2.1-μm collector junction depth, and emitter and collector widths of 22 and 8 μm, respectively. The doping profiles, contacts, and mesh were also unchanged during pre- and post-irradiation simulations. See Table **1** in **Appendix A** for details.

The devices were exposed to a $^{60}$Co source at four dose rates: 1 and 0.1 rad($SiO_2$) $s^{-1}$, and 1 and 0.167 mrad($SiO_2$) $s^{-1}$. Irradiations from 1 rad($SiO_2$) $s^{-1}$ to 1 mrad($SiO_2$) $s^{-1}$ were conducted at the Xinjiang Technical Institute of Physics and Chemistry (XTIPC). Irradiation at the ultralow dose rate of 0.167 mrad($SiO_2$) $s^{-1}$ was conducted at Peking University. All terminals of the GLPNP BJTs were grounded during irradiation.

After each irradiation step, GS and DS measurements were performed using a Keithley 4200-SCS semiconductor parameter analyzer. For both measurements, the base, collector, and substrate were maintained at 0 V. For the DS measurements, the transistors were biased in reverse mode with the emitter voltage fixed at −1 V. The emitter current was recorded while the gate-to-base voltage was swept from 0 to −40 V in 0.2 V steps. The results are shown as the solid curves in **Fig. 4**. For the GS measurements, the transistors were biased in the forward-active mode with the emitter voltage fixed at 0.5 V. The base current was recorded while the gate-to-base voltage was swept from 5 to −40 V in 0.2 V steps. The results are shown as the solid curves in **Fig. 5**.

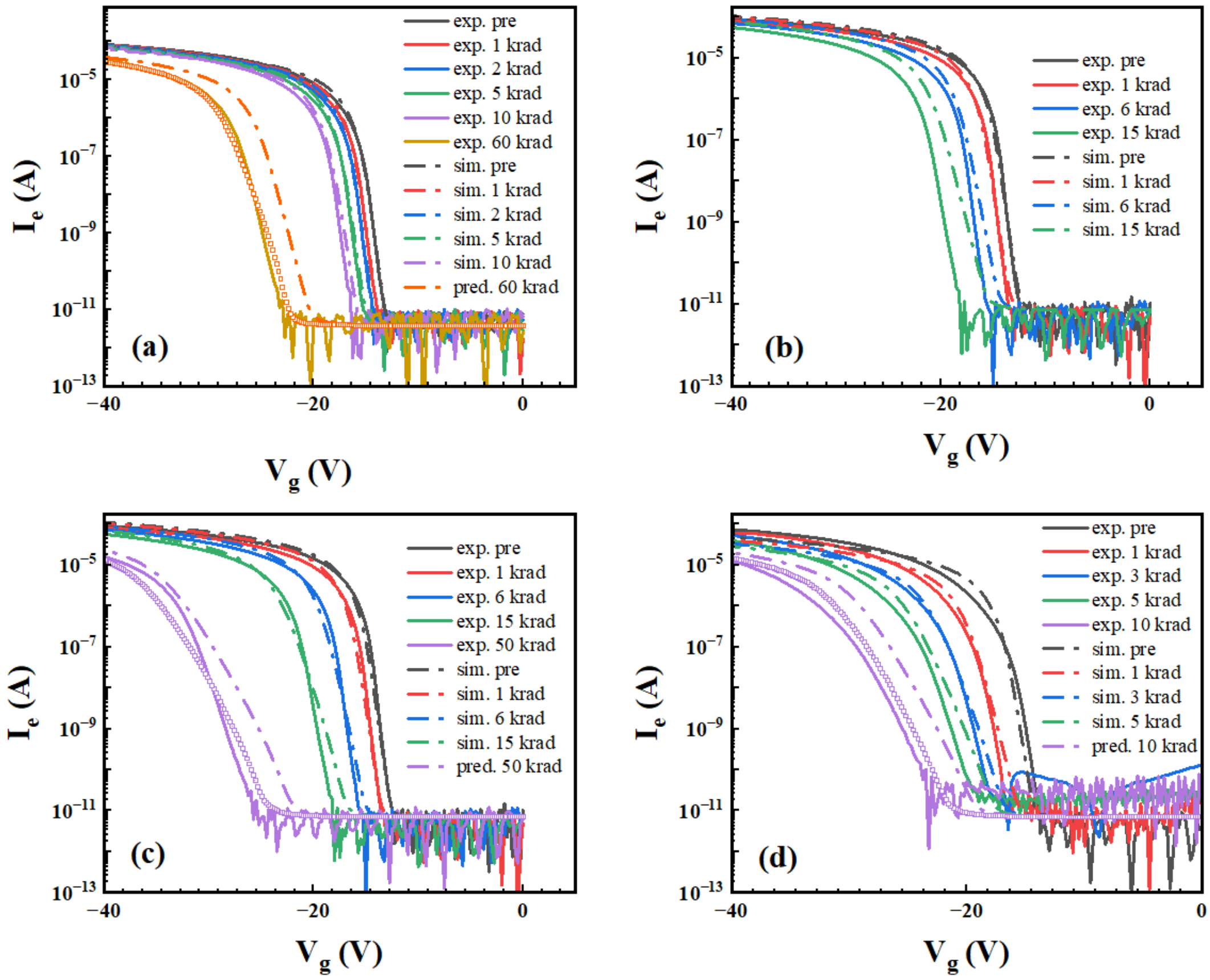

**Figure 4.** Measured (solid curves) and simulated (dashed curves) GLPNP DS curves. (a) 1 rad/s, showing the 60-krad prediction with uniform spectra (orange dash-dotted curve) and a supplementary calculation with Gaussian spectra (orange dotted curve). (b) 0.1 rad/s with uniform spectra. (c), (d) 0.1 rad/s and 1 mrad/s, respectively, with Gaussian spectra. The magenta dash-dotted curves are extrapolations to 50 and 10 krad with fixed Gaussian parameters; the magenta dotted curves include the fitted increase in donor-state width to 0.29 eV.

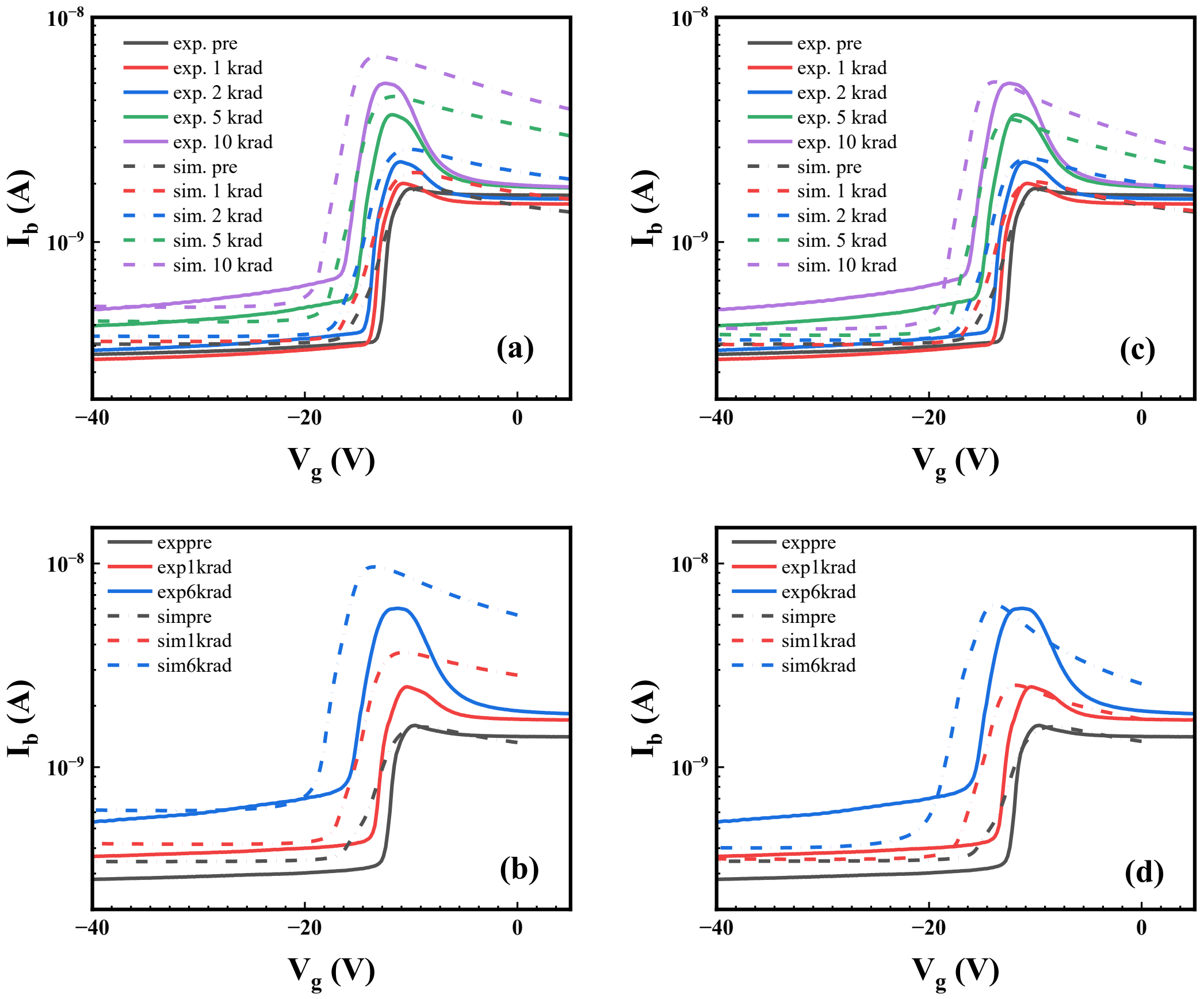

**Figure 5.** GLPNP GS curves calculated with the initial effective hole-capture cross section kept fixed at (a) 1 rad/s and (b) 0.1 rad/s, and with the adjusted effective hole-capture cross sections at (c) 1 rad/s and (d) 0.1 rad/s.

The defect concentrations are further extracted from these DS and GS curves [41], [45], [48]–[50]. The oxide capacitance per unit area was first calculated using the parallel-plate capacitor approximation, $C_{\mathrm{ox}} = \varepsilon_{\mathrm{ox}}/t_{\mathrm{ox}} = 4.60 \times 10^{-9}\ \mathrm{F/cm^2}$ , where $\varepsilon_{\mathrm{ox}}$ and $t_{\mathrm{ox}}$ are the oxide permittivity and thickness, respectively. The threshold voltage $V_{\mathrm{th}}$ was then extracted by constructing a tangent at the maximum-transconductance region of the DS curve on a linear current scale and extrapolating it to $I_E = 0$. The total defect density was calculated as $N_{\mathrm{total}} = (C_{\mathrm{ox}}/q)\Delta V_{\mathrm{th}}$. Subsequently, the midgap voltage $V_{\mathrm{mg}}$ was determined from the peak position of the GS curve, and the oxide-trap density was obtained as $N_{\mathrm{ot}} = (C_{\mathrm{ox}}/q)\Delta V_{\mathrm{mg}}$. Finally, the interface-trap density was calculated as $N_{\mathrm{it}} = N_{\mathrm{total}} - N_{\mathrm{ot}}$. The results are shown as the dots in **Fig. 6**.

### 3.2. Predictive simulation and experimental verification of DS curves at fitted and extrapolated doses

Our predictive TCAD simulation of DS curves in the dose domain is carried out following the workflow in Fig. 1(b). For any dose rate, the device is first calibrated against the pre-irradiation DS curves [dashed black in **Fig. 4**]. In these cases, the uniform interface-state energy distribution [dashed in **Fig. 7**] and capture cross sections [the pristine values in Table 2, **Appendix B**] are adopted, as suggested in [10]. In latter post-irradiation DS simulation, they are kept fixed, and only $N_{ot}$(D) and $N_{it}$(D) change with dose, defining the concentration-only simulation.

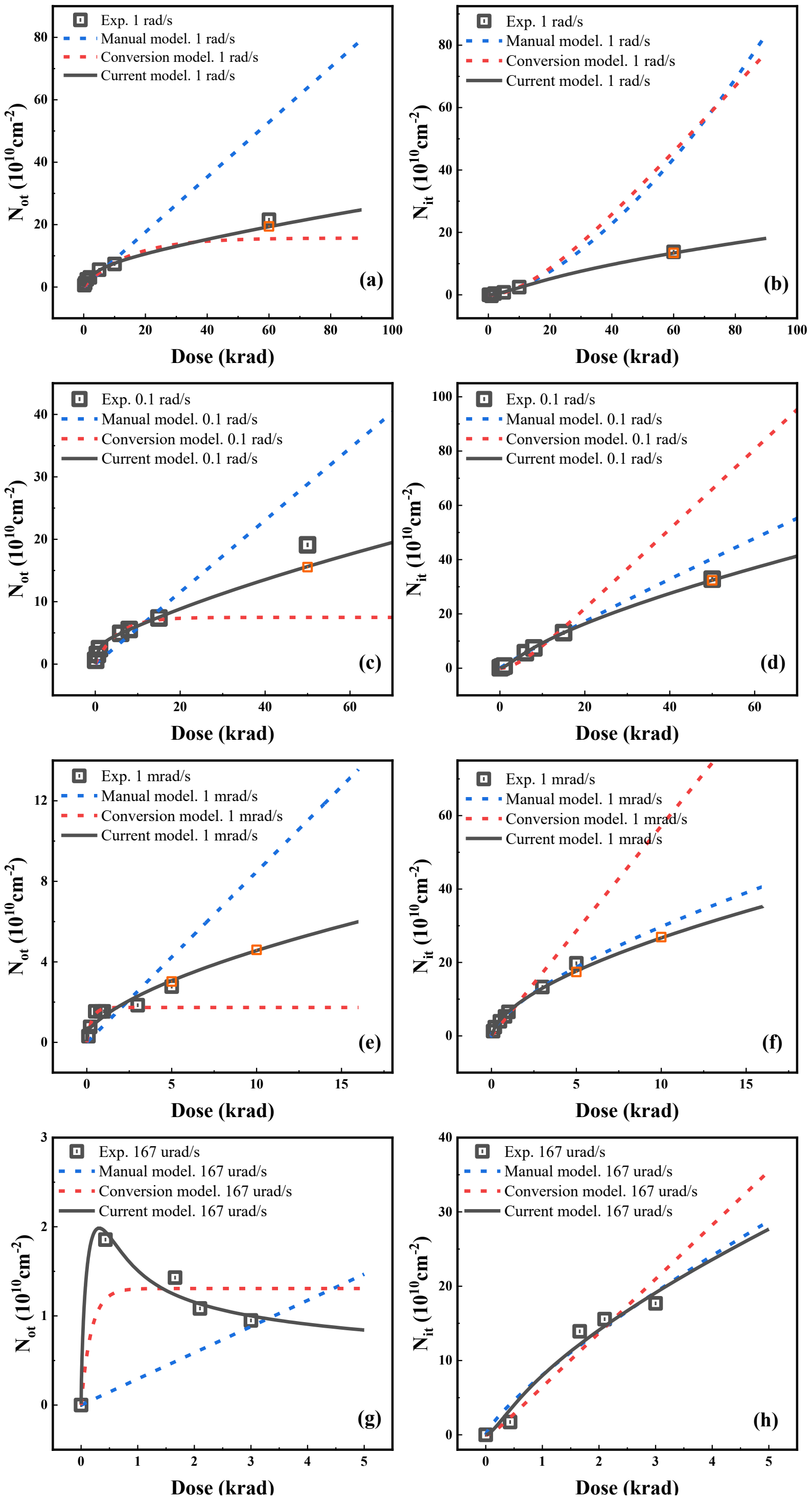

**Figure 6.** Comparison of defect models fitted to the same data: (a), (c), (e), and (g) $N_{ot}$; (b), (d), (f), and (h) $N_{it}$. (a)–(b) 1 rad/s, with 0–10-krad fitted and the 60-krad data reserved for validation. (c)–(d) 0.1 rad/s, with 0–15-krad fitted and the 50-krad data reserved. (e)–(f) 1 mrad/s, with 0–5-krad fitted and the 10-krad data reserved. (g)–(h) 0.167 mrad/s, showing the nonmonotonic regime. Black points denote measurements; orange points denote values predicted by the partially coupled model.

We first consider the relatively high dose rate of 1rad/s. As shown in Figs. **6(a)** and **(b)**, the 1–10-krad $N_{ot}$ and $N_{it}$ data are jointly fitted to Eq. (2). Accordingly, the four kinetic parameters are obtained as the red points at 1 rad/s in **Fig. 8**. Python then calculates the defect concentrations for any require total dose and transfers them to SWB for corresponding I－V simulation. As shown by the dashed red, blue, green, and purple curves in **Fig. 4(a)**, the module simultaneously and quantitatively reproduces the measured DS curves at multi doses in the fitted dose range. The agreement includes both the lateral shifts and the changes in the subthreshold region. Such simultaneous and quantitative reproduction of I − V curves at multiple doses is different from the usual dose-by-dose calibration.

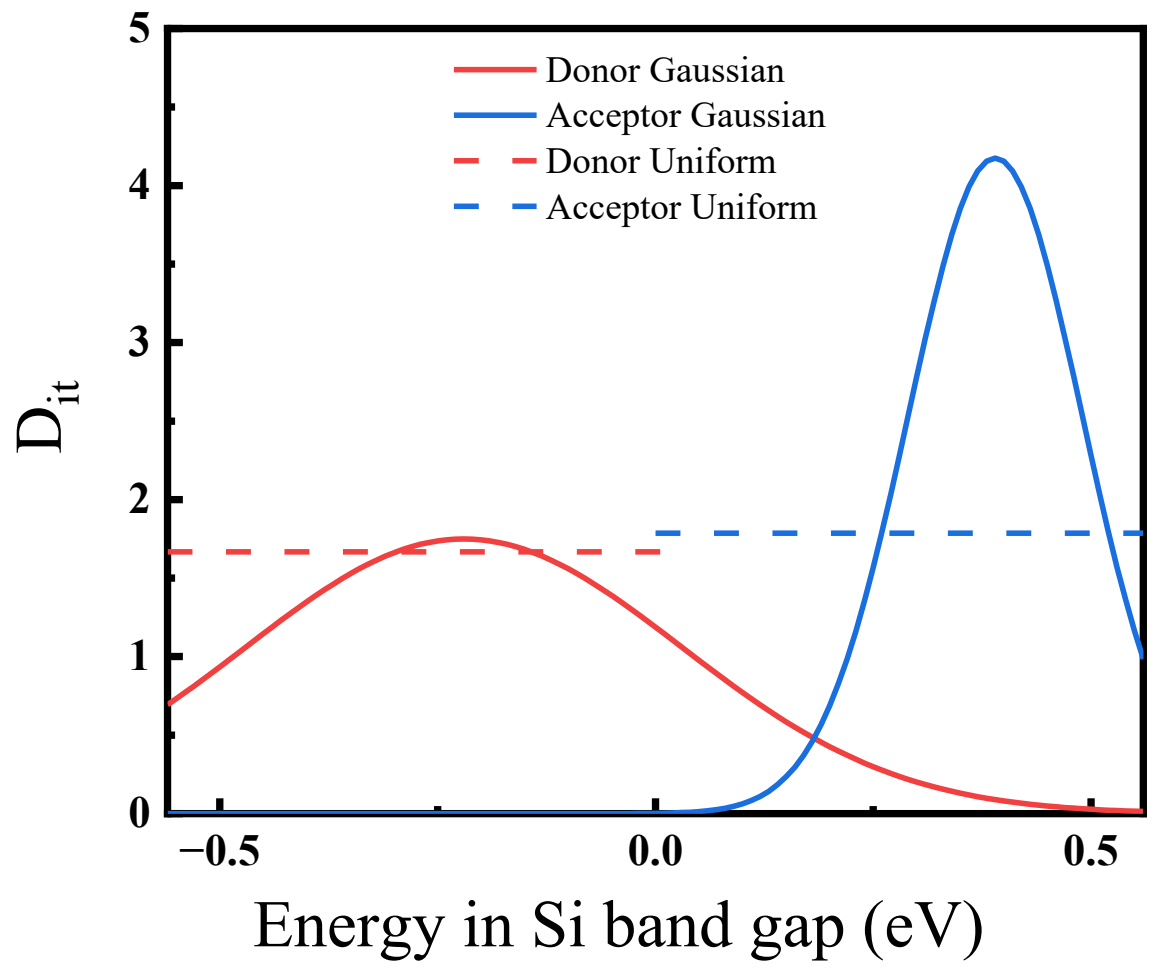

**Figure 7.** Uniform and Gaussian interface-state energy distributions (normalized) used for the DS calculations. The Gaussian center energies are 0.33 eV for donor-like states and 0.95 eV for acceptor-like states, measured from the Si valence-band edge. Their widths are 0.25 and 0.10 eV, respectively. The integrated trap densities are retained when replacing the uniform spectra with Gaussian spectra.

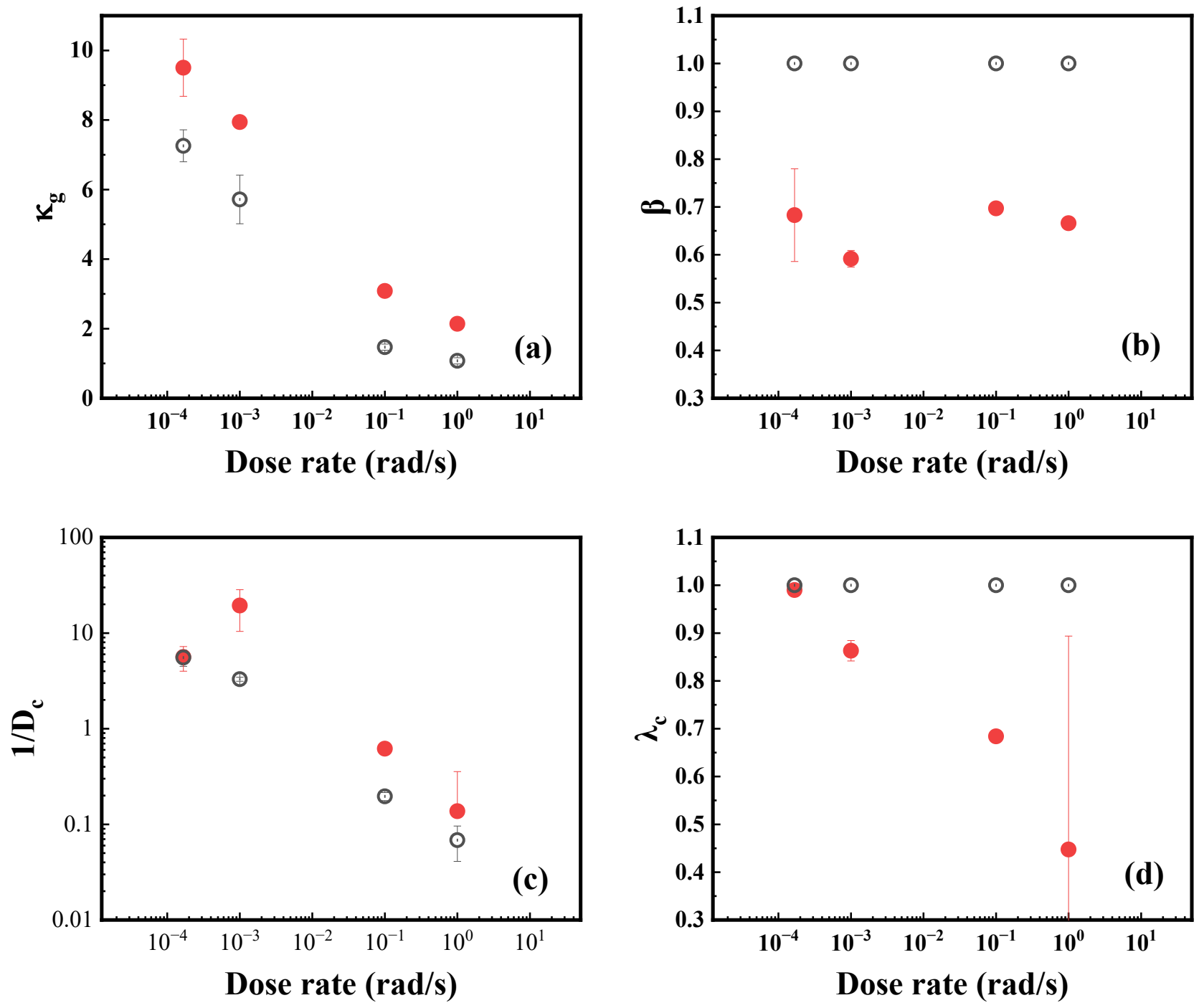

**Figure 8.** Kinetic parameters fitted at each dose rate for the partially coupled (red) and fully coupled (black) models: (a) $\kappa_g$, (b) $\beta$, (c) $1/D_c$, and (d) $\lambda_c$.

To further test the extrapolation capability of our module, the defect concentrations at 60 krad [$N_{ot}(60)$ and $N_{it}(60)$] are extrapolated using Eq. (2) with the fitted parameter set. The values are very close to the measurements [Figs. 6(a,b)]. Then, the full 60-krad DS curve is calculated following the workflow in Fig. 1(b). As shown by the orange dash-dotted curve in **Fig. 4(a)**, the predicted curve is close to the reserved measurement. The signed relative threshold-voltage error, defined as (predicted − measured)/measured, is −14%, although the tested dose is 6 times the upper fitting dose. Both threshold voltages are extracted using the method described in section 3.1. This independent comparison validates extrapolation beyond the fitted dose range. For this prediction, neither the 60-krad concentrations nor the measured I–V curve at that dose was used in any kinetic, device, energy-distribution, or capture calibration.

The result shows that with the TCAD module, smaller-dose concentration measurements can support prediction of a larger-dose electrical response, offering a way to reduce irradiation time.

## 4. Additional trap information identified by the module

### 4.1. Energy distributions for lower-dose-rate DS curves

We next apply the concentration-only procedure at lower dose rates, fitting the defect concentration data and obtaining kinetic parameters separately for each dose-rate condition [Figs**. 6(c-f)** and **Fig. 8**]. At 0.1 rad/s, the calculated $N_{ot}$(D) and $N_{it}$(D), combined with uniform interface-state energy distributions, reproduce the direction of the DS shifts but not the curve shapes at 6 and 15 krad [**Fig. 4(b)**]. The discrepancy increases with dose. Since the calculated concentrations are very closed to the defect data [Fig. 6(c,d)], we examine whether a nonuniform energy distribution can account for the remaining mismatch at relatively low dose rate.

The uniform spectra (dashed) used so far are compared with Gaussian spectra (solid) in Fig. 7. Redistributing a fixed integrated $N_{it}$ over energy changes trap occupancy and net interface charge as the gate voltage varies, which can affect both the DS subthreshold shape and the curve shift. We therefore replace the uniform acceptor and donor distributions with Gaussian functions. For acceptor-like states, $E_{\mathrm{Mid}}$ = 0.95 eV and $E_{\mathrm{Sig}}$ = 0.10 eV; for donor-like states, $E_{\mathrm{Mid}}$ = 0.33 eV and $E_{\mathrm{Sig}}$ = 0.25 eV. Both center energies are measured from the Si valence-band edge. The integrated trap densities retain the values used in the concentration-only calculations. As seen in **Fig. 4(c)**, these Gaussian spectra substantially reduce the mismatch at 6 and 15 krad for 0.1 rad/s.

We then applied the similar simulation process to a lower dose rate of 1 mrad/s. The concentration fits are shown in Figs**. 6(e)** and **(f)**, while the DS results are displayed in **Fig. 4(d)**. It is seen that, our module can reproduce the 1, 3, and 5 krad curves. These results of 0.1 rad/s and 1 mrad/s show that the energy distribution must also be specified to reproduce the lower-dose-rate DS curve shapes.

Can non-uniform distribution improve extrapolation predictions? To test this assumption, we apply the same Gaussian spectra in a supplementary 60-krad calculation at 1 rad/s. It is seen that, this further improves agreement with the measurement [orange dotted curve in **Fig. 4(a)**].

We then extrapolate the concentrations [Figs. 6(c-f)] to predict the 50-krad DS curve at 0.1 rad/s and the 10-krad curve at 1 mrad/s [magenta dash-dotted curves in Figs. 4 (c,d)]. Systematic differences from the measurements (magenta solid curves) remain. Previous experiments have reported spectral redistribution after irradiation [37], [38], [51], while gate-controlled bipolar measurements have linked states between flatband and threshold to base-current broadening [39], [40]. We therefore test whether additional spectral broadening can account for the residual mismatch. As seen by the magenta dotted curves in **Figs. 4(c**) and (**d),** increasing the donor-state $E_{\mathrm{Sig}}$ from 0.25 to 0.29 eV, while retaining the center energies, acceptor width, and integrated trap densities, substantially improves the two high-dose curves. These results further identify the important role of broadening distribution in the quantitative DS simulation. It should be noticed that, this additional width describes an effective spectral correction and does not uniquely identify the microscopic trap distribution.

### 4.2. Effective capture cross sections required for GS curves

Finally, we moved on to the simulation of GS curves, which provide a more demanding test than the DS curves because the base-current peak depends on both surface electrostatics and interface-state-assisted recombination. Even when the total interface-trap density, energy distribution, and occupancy are specified, the GS peak amplitude and width also depend on the electron and hole capture cross sections [39], [40]. Recent GLPNP experiments by Duan et al. indicate that interface-trap capture properties depend on both total dose and dose rate [52].

Motivated by these findings, we introduce a dose-dependent effective hole-capture cross section, $\sigma_{\mathrm{h,eff}}$(D), to describe the GS response at each dose rate. For the 1- and 0.1-rad/s calculations, the electron capture cross section is fixed at $10^{-15}$ cm$^2$ [**Tables 2 and 3 in Appendix B**]. The fitted $\sigma_{\mathrm{h,eff}}$(D) represents the combined recombination response of the interface-trap population. At 1 and 0.1 rad/s, the same initial $\sigma_{\mathrm{h,eff}}$ is assigned to donor- and acceptor-like interface states. Its value, $8 \times 10^{-12}$ cm$^2$, is calibrated against the pre-irradiation curves. **Figs. 5(a) and (b)** show that, keeping this value fixed after irradiation gives pronounced discrepancies between the simulations and measurements. **Figures 5(c) and (d)** show that the GS curves are well reproduced when $\sigma_{\mathrm{h,eff}}$ decreases to $2.2 \times 10^{-12}$ cm$^2$ at 10 krad for 1 rad/s and to $5 \times 10^{-13}$ cm$^2$ at 6 krad for 0.1 rad/s [see Tables 2 and 3 for details]. These results show that dose-dependent effective capture cross sections are needed to reproduce these GS curves.

The simulation of the lowest dose rate of 0.167 mrad/s test how the defect concentrations affect the GS peak position. The corresponding concentration fits are shown in Figs**. 6(g)** and **(h)**. As seen by the solid curves in **Fig. 9**, the measured peak first shifts to the left and then reverses direction slightly as dose increases. A TCAD simulation with fixed effective capture cross sections [Table **4 in Appendix B**] changes the peak magnitude and width but retains this reversal of peak positions [dashed curves in **Fig. 9**]. These results show that the peak position is governed mainly by $N_{ot}$(D), while the energy distribution and capture cross sections affect the recombination-sensitive shape and amplitude. The defect kinetics model in Eq. (2) reproduces the reversal because $N_{ot}$(D) is itself nonmonotonic: oxide-trap generation slows as the irradiation goes on, while $H_2$-mediated passivation continues.

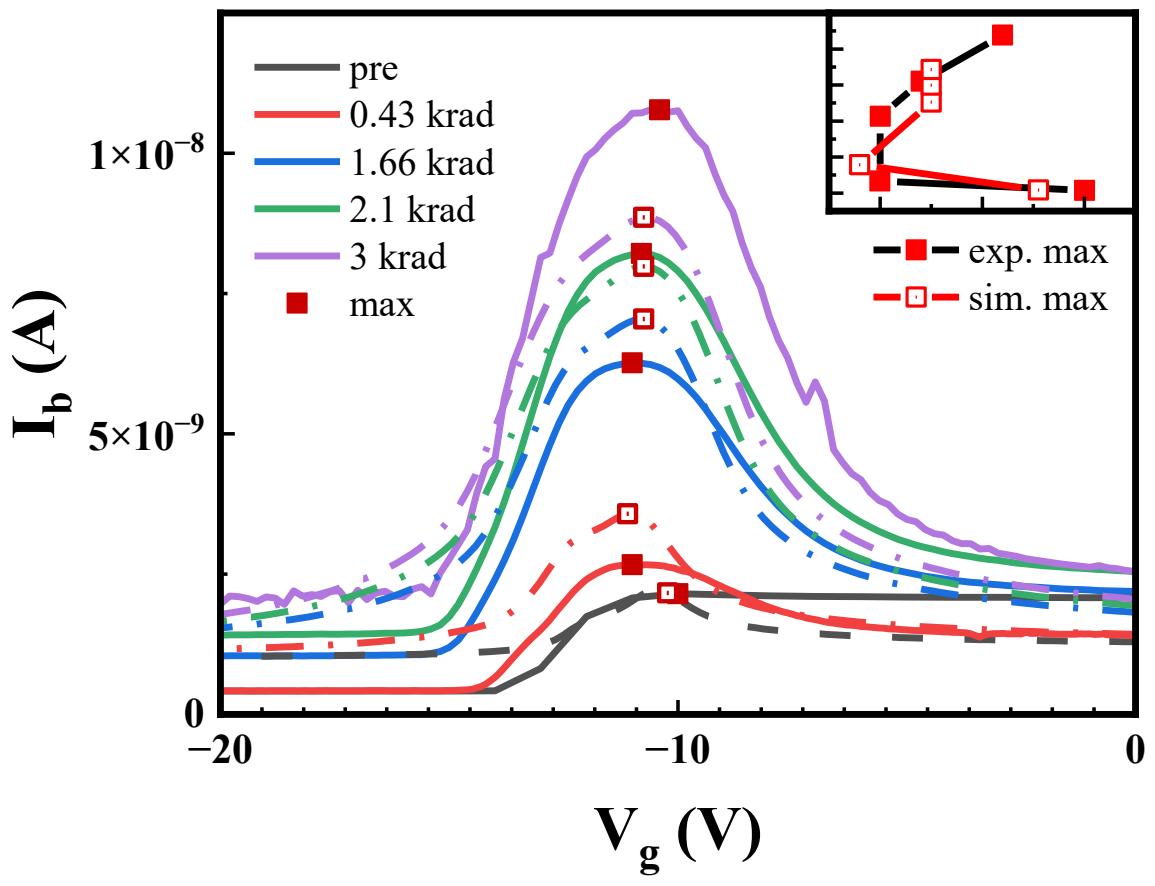

**Figure 9.** Measured (solid) and simulated (dashed) GLPNP GS curves at 0.167 mrad/s. The calculation with fixed effective capture cross sections retains the reversal in peak position (enlarged in the inset), supporting the dominant role of $N_{ot}$ in this shift.

The above comparisons identify which trap properties must be included for the quantitative simulation of electrical responses examined here. Varying concentrations alone reproduces the DS curves at 1 rad/s. The lower-dose-rate DS shapes require changes in the energy distribution, and the GS curves additionally require effective capture cross sections that vary with dose and dose rate. These findings are consistent with previous observations of spectral redistribution and capture asymmetry [37]–[40], [52], [53]. Here, their role in TCAD simulation is established by using the concentration model as a baseline.

## 5. Discussion

### 5.1. Comparison of uncoupled and coupled defect models

To demonstrate why the *partially coupled* model is needed for predictive TCAD in the dose domain, the data in **Fig. 6** are also fitted by two other defect kinetics models: (i) the *uncoupled* linear and sublinear dose laws in Eq. (1), with their coefficients fitted to the data rather than fixed; (ii) the *fully coupled* $N_{ot}$–$N_{it}$ conversion model [47].

For ease of comparison, we rewrite the fully coupled model [47] as

$$N_{ot}^{P}(D) = \kappa_g D_c[1 - exp(-D/D_c)] \quad (3a)$$

$$N_{it}^{P}(D) = \kappa_g D - N_{ot}^{P}(D), \quad (3b)$$

where $\kappa_g$ is a linear generation coefficient and $D_c = q\tau_c$ is a characteristic conversion dose. It is interesting to noticed that, this model can be recovered from Eq. (2) by setting $\beta = 1$ and $\lambda_c = 1$. These values impose linear generation and allow all generated oxide traps to undergo the passivation process associated with interface-trap formation, respectively.

The fitting results are shown as the blue and red curves in Fig. 6. It is seen that, the uncoupled model in **Eq. (1)** misses both the sublinear $N_{ot}$ buildup at higher dose rates [blue curves in Figs**. 6(a)**, **(c),** and **(e)**] and the non-monotonic response at the ultralow dose rate [blue curve in **Fig. 6(g)**]. On the other hand, the fully coupled model drives $N_{ot}$ toward saturation and $N_{it}$ toward an almost linear dose dependence [red curves in **Fig. 6**]. The saturation prediction deviates from the measured sublinear and non-monotonic data, while the near-linear behavior deviates from the fractional-power-law response of $N_{it}$ observed experimentally. In sharp contrast, across four dose-rate conditions spanning approximately four decades, the partially coupled model reproduces both the sublinear and nonmonotonic $N_{ot}$ trends and the approximately fractional-power-law buildup of $N_{it}$ [black curves]. This general fit supports the DS reconstruction at multiple doses and higher-dose extrapolation at 1 rad/s, and provides a concentration baseline for identifying the additional trap information required at lower dose rates.

Within the partially coupled model, the fitted parameters in **Figure 8** explain the different dose dependences at high and low dose rates. At the lowest dose rate, $\lambda_c$ approaches unity, indicating that nearly all generated oxide traps can participate in passivation associated with interface-trap formation. At higher dose rates, the smaller $\lambda_c$ means that a smaller fraction of generated oxide traps participates in this process. In this case, the remaining fraction of $N_{ot}$ contributes the first, monotonic generation term in Eq. (2a), giving the sublinear $N_{ot}$ buildup. At the ultralow dose rate, oxide-trap generation slows while passivation continues. With a larger $\lambda_c$, this produces the nonmonotonic $N_{ot}$ response. Meanwhile, $\beta$ remains below unity across all conditions. Both terms in Eq. (2b) inherit this dispersive generation, so $N_{it}$ retains an approximately fractional-power-law dose dependence. The fully coupled model fixes $\beta = \lambda_c = 1$, so fitting $\kappa_g$ and $D_c$ cannot recover retarded generation or a passivated fraction below unity. Together with the NMOS benchmark, these comparisons show why the partially coupled model matters for predictive device simulation.

### 5.2. Scope of dose extrapolation

The prediction demonstrated here is an extrapolation in the dose domain, i.e., from smaller to larger total dose under fixed irradiation conditions. The kinetic parameters for defect generation and evolution are fitted separately for each sample and dose rate. Therefore, predicting a low-dose-rate response from high-dose-rate data requires an additional model of their dose-rate dependence. The trends in Fig. 8 provide a basis for developing such models in future work. In addition, the energy-width parameter $E_{Sig}(D)$ and effective capture cross section $\sigma_{h,eff}(D)$ are also calibrated against the electrical data at lower dose rates. Quantifying the evolution of them as functions of dose and dose rate is a further step toward extending the present module to prediction across dose rates.

The module may also provide a TID baseline for studying synergistic effects of ionizing and displacement damage (DD) [54]–[57]. Interpreting these effects at the device level requires reliable descriptions of the separate TID and DD responses. A predictive TID baseline could help distinguish changes in carrier distributions around displacement defects caused by oxide trapped charge [58,59] from ionization-induced evolution of the displacement defects themselves [55,56].

## 6. Conclusion

We have developed a predictive TID TCAD module in the dose domain by integrating our previously established partially coupled defect kinetics model for $N_{\mathrm{ot}}(\mathrm{D})$ and $N_{\mathrm{it}}(\mathrm{D})$ [21] into Sentaurus. Two advances have been achieved. First, at 1 rad/s, one parameter set fitted to the 1–10-krad concentrations reproduces the GLPNP DS curves and predicts the full 60-krad curve. This demonstrates extrapolation from smaller to larger total dose under fixed irradiation conditions, which can reduce irradiation time. Second, calculations constrained by the model concentrations further identify the additional information required for other responses: lower-dose-rate DS curves need nonuniform, dose-dependent interface-state energy distributions, modeled here with Gaussian functions, and the GS curves at 1 and 0.1 rad/s additionally need dose- and dose-rate-dependent effective capture cross sections. Therefore, our TCAD module supports predictive TID simulation for radiation qualification.

## Funding sources

This work was supported by the Xinjiang Tianchi Talents Program (Grant No. 2025000061), the Distinguished Young Scholars Project of the Natural Science Foundation of Xinjiang Uygur Autonomous Region (Grant No. 2027D01E58), the National Natural Science Foundation of China (Grant No. U2530216), the Autonomous Region Natural Science Foundation Key Project under Grant No. 2024D01D30, the State Key Laboratory of Semiconductor Physics and Chip Technologies, Institute of Semiconductors, Chinese Academy of Sciences (Grant No. SKSP202507), and the fund of the Innovation Center for Radiation Application (Grant No. KFZC2026020501).

## Acknowledgements

Y. Song thanks Dr. Hang Zhou and Dr. Guanghui Zhang for insightful discussions.

## Appendix

### A. GLPNP Geometry and Doping Parameters

**Table 1.** Geometry and doping parameters used in the GLPNP TCAD model

| Category | Parameter | Value |
|---|---|---|
| **Geometry constants** | Oxide thickness | 0.75 μm |
| **Geometry constants** | Base width | 14.0 μm |
| **Geometry constants** | Base–collector lateral spacing | 22.0 μm |
| **Geometry constants** | N-epi / base vertical thickness | 12.0 μm |
| **Geometry constants** | Collector junction depth | 2.1 μm |
| **Geometry constants** | Collector width | 8.0 μm |
| **Geometry constants** | Emitter width | 22.0 μm |
| **Doping profiles** | P-Sub constant profile | Boron Active Concentration = $1\times10^{14}$ $cm^{-3}$ |
| **Doping profiles** | N-epi constant profile | Phosphorus Active Concentration = $1\times10^{15}$ $cm^{-3}$ |
| **Doping profiles** | BN Gaussian reference line | x = 0–162 μm， y = 24 μm |
| **Doping profiles** | BN1 | Phosphorus； peak $1\times10^{19}$； at 4 μm: $4\times10^{16}$ $cm^{-3}$； factor 0.2； Negative |
| **Doping profiles** | BN2 | Phosphorus； peak $1\times10^{19}$； at 4 μm: $9\times10^{14}$ $cm^{-3}$； factor 0.2； Positive |
| **Doping profiles** | Base region | Phosphorus； peak $2\times10^{20}$； at 12 μm: $1\times10^{17}$ $cm^{-3}$； factor 0.2； Negative |
| **Doping profiles** | Collector region | Boron； peak $1\times10^{18}$； at 2.1 μm: $1\times10^{15}$ $cm^{-3}$； factor 0.5； Negative |
| **Doping profiles** | Emitter region | Boron； peak $1\times10^{18}$； at 2.1 μm: $1\times10^{15}$ $cm^{-3}$； factor 0.5； Negative |

## B. Dose- and Dose-Rate-Dependent Interface-State Parameters

**Table 2.** Interface-state parameters for GS simulations at 1 rad/s

| Dose | State type | $\sigma_e$(cm$^2$) | $\sigma_h$(cm$^2$) |
|---|---|---|---|
| Pristine | Acceptor | $1\times10^{-15}$ | $8\times10^{-12}$ |
| | Donor | $1\times10^{-15}$ | $8\times10^{-12}$ |
| 1 krad | Acceptor | $1\times10^{-15}$ | $6\times10^{-12}$ |
| | Donor | $1\times10^{-15}$ | $6\times10^{-12}$ |
| 2 krad | Acceptor | $1\times10^{-15}$ | $6\times10^{-12}$ |
| | Donor | $1\times10^{-15}$ | $6\times10^{-12}$ |
| 5 krad | Acceptor | $1\times10^{-15}$ | $3.4\times10^{-12}$ |
| | Donor | $1\times10^{-15}$ | $3.4\times10^{-12}$ |
| 10 krad | Acceptor | $1\times10^{-15}$ | $2.2\times10^{-12}$ |
| | Donor | $1\times10^{-15}$ | $2.2\times10^{-12}$ |

**Table 3**. Interface-state parameters for GS simulations at 0.1 rad/s

| Dose | State type | $\sigma_e$(cm$^2$) | $\sigma_h$(cm$^2$) |
|---|---|---|---|
| Pristine | Acceptor | $1\times10^{-15}$ | $8\times10^{-12}$ |
| | Donor | $1\times10^{-15}$ | $8\times10^{-12}$ |
| 1 krad | Acceptor | $1\times10^{-15}$ | $9\times10^{-13}$ |
| | Donor | $1\times10^{-15}$ | $9\times10^{-13}$ |
| 6 krad | Acceptor | $1\times10^{-15}$ | $5\times10^{-13}$ |
| | Donor | $1\times10^{-15}$ | $5\times10^{-13}$ |

**Table 4.** Interface-state parameters for GS simulations at 0.167 mrad/s

| Dose | State type | $\sigma_e$(cm$^2$) | $\sigma_h$(cm$^2$) |
|---|---|---|---|
| All | Acceptor | $1\times10^{-14}$ | $1\times10^{-15}$ |
| | Donor | $1\times10^{-15}$ | $1\times10^{-14}$ |